# 21st Century Computer Architecture

*A community white paper*

*May 25, 2012*

## 1. Introduction and Summary

Information and communication technology (ICT) is transforming our world, including healthcare, education, science, commerce, government, defense, and entertainment. It is hard to remember that 20 years ago the first step in information search involved a trip to the library, 10 years ago social networks were mostly physical, and 5 years ago "tweets" came from cartoon characters.

Importantly, much evidence suggests that ICT innovation is accelerating with many compelling visions moving from science fiction toward reality[1]. Appendix A both touches upon these visions and seeks to distill their attributes. Future visions include personalized medicine to target care and drugs to an individual, sophisticated social network analysis of potential terrorist threats to aid homeland security, and telepresence to reduce the greenhouse gases spent on commuting. Future applications will increasingly require processing on large, heterogeneous data sets ("Big Data"[2]), using distributed designs, working within form-factor constraints, and reconciling rapid deployment with efficient operation.

Two key—but often invisible—enablers for past ICT innovation have been semiconductor technology and computer architecture. Semiconductor innovation has repeatedly provided more transistors (Moore's Law) for roughly constant power and cost per chip (Dennard Scaling). Computer architects took these rapid transistor budget increases and discovered innovative techniques to scale processor performance and mitigate memory system losses. The combined effect of technology and architecture has provided ICT innovators with exponential performance growth at near constant cost.

Because most technology and computer architecture innovations were (intentionally) invisible to higher layers, application and other software developers could reap the benefits of this progress without engaging in it. Higher performance has both made more computationally demanding applications feasible (e.g., virtual assistants, computer vision) and made less demanding applications easier to develop by enabling higher-level programming abstractions (e.g., scripting languages and reusable components). Improvements in computer system cost-effectiveness enabled value creation that could never have been imagined by the field's founders (e.g., distributed web search sufficiently inexpensive so as to be covered by advertising links).

---

[1] PCAST, "Designing a Digital Future: Federally Funded Research and Development Networking and Information Technology, Dec. 2010 (http://www.whitehouse.gov/sites/default/files/microsites/ostp/pcast-nitrd-report-2010.pdf).

[2] CCC, "Challenges and Opportunities with Big Data," Feb. 2012 (http://cra.org/ccc/docs/init/bigdatawhitepaper.pdf).

The wide benefits of computer performance growth are clear. Recently, Danowitz et al.[3] apportioned computer performance growth roughly equally between technology and architecture, with architecture credited with ~80× improvement since 1985. As semiconductor technology approaches its "end-of-the-road" (see below), computer architecture will need to play an increasing role in enabling future ICT innovation. But instead of asking, "How can I make my chip run faster?," architects must now ask, "**How can I enable the 21st century infrastructure, from sensors to clouds, adding value from performance to privacy, but without the benefit of near-perfect technology scaling?**". The challenges are many, but with appropriate investment, opportunities abound. Underlying these opportunities is a common theme that future architecture innovations will require the engagement of and investments from innovators in other ICT layers.

## 1.1 The Challenges: An Inflection Point

The semiconductor technology enabler to ICT is facing serious challenges that are outlined in Table 1 below. First, although technologists can make more and smaller transistors (Moore's Law), these transistors are not altogether "better" as has been true for four decades. Second, the power per transistor is no longer scaling well (Dennard Scaling has ended). Since most products—sensors, mobile, client, and data center—cannot tolerate (repeated) power increases, we must consider ways to mitigate these increases. Third, fabrication variations of nano-scale features (e.g., gate oxides only atoms thick) reduce transistors' long-term reliability significantly compared to larger feature sizes. Fourth, communication among computation elements must be managed through locality to achieve goals at acceptable cost and energy with new opportunities (e.g., chip stacking) and new challenges (e.g., data centers). Fifth, one-time costs to design, verify, fabricate, and test are growing, making them harder to amortize, especially when seeking high efficiency through platform specialization (e.g., handhelds, laptops, or servers).

**Table 1: Technology's Challenges to Computer Architecture**

| Late 20th Century | The New Reality |
|---|---|
| Moore's Law — 2× transistors/chip every 18-24 months | Transistor count still 2× every 18-24 months, but see below |
| Dennard Scaling — near-constant power/chip | Gone. Not viable for power/chip to double (with 2× transistors/chip growth) |
| The modest levels of transistor unreliability easily hidden (e.g., via ECC) | Transistor reliability worsening, no longer easy to hide |
| Focus on computation over communication | Restricted inter-chip, inter-device, inter-machine communication (e.g. Rent's Rule, 3G, GigE); communication more expensive than computation |
| One-time (non-recurring engineering) costs growing, but amortizable for mass-market parts | Expensive to design, verify, fabricate, and test, especially for specialized-market platforms |

---

[3] Danowitz, et al., "CPU DB: Recording Microprocessor History", CACM 04/2012.

# 1.2 The Opportunities: 21st Century Computer Architecture

With CMOS technology scaling weakening as an enabler of ICT innovation, computer architects must step up their role even further. 21st century computer architecture, however, needs to be different from its 20th century predecessor to embrace this new role. We see three fundamental differences, highlighted in Table 2 below. These differences form the basis of the future research agenda described in Section 2.

**Table 2: Computer Architecture's Past and Future**

| 20th Century Architecture | 21st Century Architecture | |
|---|---|---|
| Single-chip performance | ***Architecture as infrastructure:*** *from sensors to clouds* <br> ● Chips to systems <br> ● Performance plus security, privacy, availability, programmability, … | ***Cross-cutting implication:*** <br><br> Break current layers with new interfaces |
| Performance through software-invisible instruction level parallelism (ILP) | ***Energy first*** <br> ● Parallelism <br> ● Specialization <br> ● Cross-layer design | |
| Tried and tested technologies: CMOS, DRAM, disks with rapid but predictable improvements | ***New technologies:*** *non-volatile memory, near-threshold voltage operation, 3D chips, photonics, …* <br> Rethink <br> ● Memory+storage <br> ● Reliability <br> ● Communication <br> ● … | |

**Architecture as Infrastructure: From Sensors to Clouds**: Past architecture research often focused on a chip (microprocessor) or stand-alone computer with performance as its main optimization goal. Moving forward, computers will be a key pillar of the 21st century societal infrastructure. To address this change, computer architecture research must expand to recognize that generic computers have been replaced by computation in context (e.g., sensor, mobile, client, data center) and many computer systems are large and geographically distributed.[4] This shift requires more emphasis on the system (e.g., communication becomes a full-fledged partner of computation), the driver application (e.g., dealing with big data), and human-centric design goals beyond performance (e.g., programmability, privacy, security, availability, battery life, form factor).

**Energy First:** Past computer architecture optimized performance, largely through software invisible changes. 21st century architecture confronts power and energy as the dominant constraints, and can no longer sustain the luxury of software invisible innovation. We see parallelism, specialization, and cross-layer design as key principles in an energy-first era, but all three require addressing significant challenges. For example, while parallelism will abound in future applications (big data = big parallelism), communication energy will outgrow computation

---
[4] Luiz André Barroso and Urs Hölzle, "The Datacenter as a Computer", Morgan-Claypool, 2009

energy and will require rethinking how we design for 1,000-way parallelism. Specialization can give 100× higher energy efficiency than a general-purpose compute or memory unit, but no known solutions exist today for harnessing its benefits for broad classes of applications cost-effectively. Cross-layer design (from circuit to architecture to run-time system to compiler to application) can wring out waste in the different layers for energy efficiency, but needs a far more concerted effort among many layers to be practical.

**New Technologies:** Past computer architecture has relied on the predictable performance improvements of stable technologies such as CMOS, DRAM, and disks. For the first time in the careers of many ICT professionals, new technologies are emerging to challenge the dominance of the "tried-and-tested," but sorely need architectural innovation to realize their full potential. For example, non-volatile memory technologies (e.g., Flash and phase change memory) drive a rethinking of the relationship between memory and storage. Near-threshold voltage operation has tremendous potential to reduce power but at the cost of reliability, driving a new discipline of resiliency-centered design. Photonics and 3D chip stacking change communication costs radically enough to affect the entire system design.

Underlying all of the above is a **cross-cutting** theme of innovations that are exposed to and require interaction with other ICT layers. This change is dramatic in that it will impact ICT innovators in other layers, similar to, but potentially greater than, the recent shift to multicore processors. Collaboration with other-layer innovators will empower architects to make bolder innovations with commensurate benefits, but it will also require significant investment and strong leadership to provide the richer inter-layer interfaces necessary for the 21$^{st}$ century.

## 2. Research Directions

This section discusses important research directions for computer architecture and related communities, but begins with two comments. First, the ideas below represent good ideas from those who contributed to this document and complement other recent documents.[5][6][7] They do not represent an exhaustive list nor a consensus opinion of the whole community, which are infeasible to gather in the process of creating this white paper.

Second, if we have been convincing that the directions presented have value to society, there is a need for pre-competitive research funding to develop them. Even highly-successful computer companies lack the incentive to do this work for several reasons. First, the technologies required will take years to a decade to develop. Few companies have such staying power. Second, successful work will benefit many companies, disincentivizing any one to pay for it. Third, many of the approaches we advocate cross layers of the system stack, transcending industry-standard interfaces (e.g., x86) and the expertise of individual companies. Finally, we need to educate the next generation of technical contributors, which is perhaps academia's most important form of technology transfer.

---

[5] ACAR-1, "Failure is not an Option: Popular Parallel Programming," Workshop on Advancing Computer Architecture Research, August 2010 (http://www.cra.org/ccc/docs/ACAR_Report_Popular-Parallel-Programming.pdf).

[6] ACAR-2, "Laying a New Foundation for IT: Computer Architecture for 2025 and Beyond," Workshop on Advancing Computer Architecture Research, September 2010 (http://www.cra.org/ccc/docs/ACAR2-Report.pdf).

[7] Fuller and Millett, "The Future of Computing Performance: Game Over or Next Level?," The National Academy Press, 2011 (http://books.nap.edu/openbook.php?record_id=12980&page=R1).

## 2.1. Architecture as Infrastructure: Spanning Sensors to Clouds

Until recently, computer systems were largely beige boxes nestled under a desk or hidden in a machine room, and computer architecture research focused primarily on the design of desktop, workstation and server CPUs. The new reality of 21st-century applications calls for a broader computer architecture agenda beyond the beige box. The emerging applications described in Appendix A demand a rich ecosystem that enables ubiquitous embedded sensing/compute devices that feed data to "cloud servers" and warehouse-scale facilities, which can process and supply information to edge devices, such as tablets and smartphones. Each class of computing device represents a unique computing environment with a specific set of challenges and opportunities, yet all three share a driving need for improved energy efficiency (performance per watt) as an engine for innovation. Moreover, future architecture research must go beyond optimizing devices in isolation, and embrace the challenges of cross-environment co-design to address the needs of emerging applications.

**Smart Sensing and Computing**. In the smart sensors space, the central requirement is to compute within very tight energy, form-factor, and cost constraints. The need for greater computational capability is driven by the importance of filtering and processing data where it is generated/collected (e.g., distinguishing a nominal biometric signal from an anomaly), because the energy required to communicate data often outweighs that of computation. This environment brings exciting new opportunities like designing systems that can leverage intermittent power (e.g., from harvested energy), extreme low-voltage (near-threshold and analog) design, new communication modalities (e.g., broadcast communication from building lighting), and new storage technologies (e.g., NVRAM). As sensors become critical to health and safety, their security and reliability must also be assured (e.g., consider recent demonstrations of remote hacking of pacemakers), including design correctness of hardware and software components. Additionally, given that sensor data is inherently approximate, it opens the potential to effectively apply approximate computing techniques, which can lead to significant energy savings (and complexity reduction).

**Portable Edge Devices**. The portable computing market has seen explosive growth, with smartphone sales recently eclipsing the PC market.[8] And yet, current devices still fall far short of the futuristic capabilities society has envisioned, from the augmented reality recently suggested by "Google Glasses," to the medical tricorder posited by Star Trek nearly 50 years ago and recently resurrected in the form of an X Prize challenge competition[9]. Enriching applications in this environment will need orders of magnitude improvement in operations/watt (from today's ~10 giga-operations/watt), since user interfaces seem to have a significant appetite for computation (e.g., multi-touch interfaces, voice recognition, graphics, holography, and 3D environment reconstruction) and even mobile applications are becoming data and compute intensive. As discussed in later sections, such systems clearly need both parallelism and specialization (the latest generation iPad's key chip has multiple cores and dedicates half of its chip area for specialized units). Given the proximity with the user, such devices motivate ideas that bring human factors to computer design, such as using user feedback to adjust voltage/frequency to save energy, focusing computation on where the user is looking, reducing the image to salient features only, or predicting and prefetching for what the user is likely to do. This environment also motivates features beyond raw performance, such as security/privacy.

---

[8] e.g., http://mashable.com/2012/02/03/smartphone-sales-overtake-pcs/ and
http://www.canalys.com/newsroom/smart-phones-overtake-client-pcs-2011
[9] http://www.xprize.org/x-prize-and-qualcomm-announce-10-million-tricorder-prize

**The Infrastructure—Cloud Servers**. Many of the most exciting emerging applications, such as simulation-driven drug discovery, or interactive analysis of massive human networks, can only be achieved with reasonable response times through the combined efforts of tens of thousands of processors acting as a single warehouse-scale computer. Internet search has demonstrated the societal importance of this computing paradigm. However, today's production search systems require massive engineering effort to create, program, and maintain, and they only scratch the surface of what might be possible (e.g., consider the potential of IBM's Watson). Systems architects must devise programming abstractions, storage systems, middleware, operating system, and virtualization layers to make it possible for conventionally trained software engineers to program warehouse-scale computers.

While many computing disciplines—operating systems, networking, and others—play a role in data center innovations, it is crucial for computer architects to consider the interface designs and hardware support that can best enable higher-level innovations. In addition, a key challenge lies in reasoning about locality and enforcing efficient locality properties in data center systems, a burden which coordination of smart tools, middleware and the architecture might alleviate. A further challenge lies in making performance predictable; as requests are parallelized over more systems, infrequent tail latencies become performance critical (if 100 systems must jointly respond to a request, 63% of requests will incur the 99-percentile delay of the individual systems due to waiting for stragglers[10]); architectural innovations can guarantee strict worst-case latency requirements. Memory and storage systems consume an increasing fraction of the total data center power budget, which one might combat with new interfaces (beyond the JEDEC standards), novel storage technologies, and 3D stacking of processors and memories.

**Putting It All Together—Eco-System Architecture.** There is a need for runtime platforms and virtualization tools that allow programs to divide effort between the portable platform and the cloud while responding dynamically to changes in the reliability and energy efficiency of the cloud uplink. How should computation be split between the nodes and cloud infrastructure? How can security properties be enforced efficiently across all environments? How can system architecture help preserve privacy by giving users more control over their data? Should we co-design compute engines and memory systems?

The research directions outlined above will push architecture research far beyond the beige box. The basic research challenges that relate all of these opportunities are improving energy efficiency dramatically and embracing new requirements such as programmability, security, privacy and resiliency from the ground up. Given the momentum in other areas (e.g., HCI, machine learning, and ubiquitous computing) now is the moment to explore them. Success will require significant investment in academic research because of the need for community-scale effort and significant infrastructure. While mobile and data centers are relatively new to the menagerie of computing devices, the research challenges we discuss will likely also apply to other devices yet to emerge.

## 2.2. Energy First

The shift from sequential to parallel (multicore) systems has helped increase performance while keeping the power dissipated per chip largely constant. Yet many current parallel computing systems are already power or energy constrained. At one extreme, high-end supercomputers and data centers require expensive many-megawatt power budgets; at the other, high-functionality sensors and portable devices are often limited by their battery's energy capacity. Portable and sensor devices typically require high performance for short periods followed by relatively long idle periods. Such bimodal usage is not typical in high-end servers, which are rarely completely idle and seldom need to operate at their maximum rate. Thus, power and energy solutions in the server space are likely to differ from those that work best in the portable device space. However, as we demand more from our computing systems—both servers and sensors—more of them will be limited by power, energy, and thermal constraints. Without new approaches to power- and energy-efficient designs and new packaging and cooling approaches, producing ICT systems capable of meeting the computing, storage and communication demands of the emerging applications described in Appendix A will likely be impossible. It is therefore urgent to invest in research to make computer systems *much* more energy efficient.

---

[10] J. Dean. "Achieving Rapid Response Times in Large Online Services." Talk in Berkeley, CA, Mar. 2012.

As the next subsections describe, energy must be reduced by attacking it across many layers, rethinking parallelism, and with effective use of specialization.

**Energy Across the Layers**

Electronic devices consume energy as they do work (and, in the case of CMOS, just by being powered on). Consequently, all of the layers of the computing stack play a role to improve energy and power efficiency: device technologies, architectures, software systems (including compilers), and applications. Therefore, we believe that a major interdisciplinary research effort will be needed to substantially improve ICT system energy efficiency. We suggest as a goal to improve the energy efficiency of computers by two-to-three orders of magnitude, to obtain, by the end of this decade, an exa-op data center that consumes no more than 10 megawatts (MW), a peta-op departmental server that consumes no more than 10 kilowatts (KW), a tera-op portable device that consumes no more than 10 watts (W), and a giga-op sensor system that consumes no more than 10 milliwatts (mW). Such an ambitious plan can only be attained with aggressive changes in all layers of the computing stack.

**At the Circuit/Technology Level**, we need research in new devices that are fundamentally more energy efficient, both in CMOS and in emerging device technologies. Research is also needed on new technologies that can improve the energy efficiency of certain functions, such as photonics for communication, 3D-stacking for integration, non-resistive memories for storage, and efficient voltage conversion. Novel circuit designs are also needed: circuits that work at ultra-low supply voltages, circuits that carry out efficient power distribution, circuits that perform aggressive power management, and circuits that are able to support multiple voltage and frequency domains on chip.

**At the Architecture Level**, we need to find more efficient, streamlined many-core architectures. We need chip organizations that are structured in heterogeneous clusters, with simple computational cores and custom, high-performance functional units that work together in concert. We need research on how to minimize communication, since energy is largely spent moving data. Especially in portable and sensor systems, it is often worth doing the computation locally to reduce the energy-expensive communication load. As a result, we also need more research on synchronization support, energy-efficient communication, and in-place computation.

**At the Software Level**, we need research that minimizes unnecessary communication. We require runtimes that manage the memory hierarchy and orchestrate fine-grain multitasking. We also need research on compilation systems and tools that manage and enhance locality. At the programming-model level, we need environments that allow expert programmers to exercise full machine control, while presenting a simple model of localities to low-skilled programmers. At the application level, we need algorithmic approaches that are energy-efficient via reduced operation count, precision, memory accesses, and interprocessor communication, and that can take advantage of heterogeneous systems. At the compiler level, we need to find ways to trade off power efficiency and performance, while also considering the reliability of the resulting binary. Overall, it is only through a fully-integrated effort that cuts across all layers that we can make revolutionary progress.

**Exploiting Parallelism to Enable Future Applications**

Throughout the 20th century, the growth in computing performance was driven primarily by single-processor improvements. But by 2004, diminishing returns from faster clock speeds and increased transistor counts resulted in serious power (and related thermal) problems. By shifting to multiple cores per chip, our industry continued to improve aggregate performance and performance per watt. But just replicating cores does not adequately address the energy and scaling challenges on chip, nor does it take advantage of the unique features and constraints of being on chip. Future growth in computer performance must come from massive on-chip parallelism with simpler, low-power cores, architected to match the kinds of fine-grained parallelism available in emerging applications.

Unfortunately, we are far from making parallel architectures and programming usable for the large majority of users. Much of the prior research has focused on coarse-grained parallelism using standard processor cores as the building block. Placing massive parallelism on a single chip offers new opportunities for parallel architecture and associated programming techniques. To unlock the potential of parallel computing in a widely-applicable form, we may need at least a decade of *concerted*, *broad-based* research to address emerging application problems at all levels of parallelism. To ensure that computing performance growth will continue to fuel new innovations, and given the magnitude of the technical challenges and the stakes involved, we need major funding investments in this area.

**Reinventing Computing Stack for Parallelism:** We recommend an ambitious, interdisciplinary effort to re-invent the classical computing stack for parallelism—programming language, compiler and programming tools, runtime, virtual machine, operating system, and architecture. Along with performance goals, energy considerations must be a first-class design constraint to forge a viable path to scalable future systems. The current layers have been optimized for uni-processor systems and act as barriers to change. Since a single, universal programming model may not exist, we recommend exploring multiple models and architectures. Moreover, different solutions are clearly needed for experts, who may interact with the innards of the machine, and the large majority of programmers, who should use simple, sequential-like models perhaps enabled by domain-specific languages.

**Applications-Focused Architecture Research:** We recommend an application-focused approach to parallel architecture research, starting from chips with few to potentially hundreds of cores, to distributed systems, networking structures at different scales, parallel memory systems, and I/O solutions. The challenge is to consider application characteristics, but without overfitting to particular specifics; we must leverage mechanisms (including programming languages) that will perform well and are power efficient on a variety of parallel systems. Fundamental architecture questions include the types of parallelism (e.g., data or task), how units should be organized (e.g., independent cores or co-processors), and how to synchronize and communicate. Solutions may differ depending on the inherent application parallelism and constraints on power, performance, or system size.

**Hardware/Software Co-Design:** Current challenges call for integrative research on parallelism that spans both software and hardware, from applications and algorithms through systems software and hardware. As discussed below, we see a need to break through existing abstractions to find novel mechanisms and policies to exploit locality and enable concurrency, effectively support synchronization, communication and scheduling, provide platforms that are programmable, high-performance, and energy efficient, and invent parallel programming models, frameworks, and systems that are truly easy to use. Given the large range of issues, we recommend a broad and inclusive research agenda.

## Enabling Specialization for Performance and Energy Efficiency

For the past two or more decades, general-purpose computers have driven the rapid advances in and society's adoption of computing. Yet the same flexibility that makes general-purpose computers applicable to most problems causes them to be energy inefficient for many emerging tasks. Special-purpose hardware accelerators, customized to a single or narrow-class of functions, can be orders of magnitude more energy-efficient by stripping out the layers of mechanisms and abstractions that provide flexibility and generality. But current success stories, from medical devices and sensor arrays to graphics processing units (GPUs), are limited to accelerating narrow classes of problems. Research is needed to (1) develop architectures that exploit both the performance and energy-efficiency of specialization while broadening the class of applicable problems and (2) reduce the non-recurring engineering (NRE) costs for software and hardware that limit the utility of customized solutions.

**Higher-level Abstractions to Enable Specialization.** General-purpose computers can be programmed in a range of higher-level languages with sophisticated tool chains that translate to fixed ISAs, providing functional and performance portability across a wide range of architectures. Special-purpose accelerators, in contrast, are frequently programmed using much lower-level languages that often directly map to hardware (e.g., Verilog), providing limited functional or performance portability and high NRE costs for software. As further discussed below, research is needed to develop new layers and abstractions that capture enough of a computation's structure to enable efficient creation of or mapping to special-purpose hardware, without placing undue burden on the programmer. Such systems will enable rapid development of accelerators by reducing or eliminating the need to retarget applications to every new hardware platform.

**Energy-Efficient Memory Hierarchies.** Memory hierarchies can both improve performance and reduce memory system energy demands, but are usually optimized for performance first. But fetching the operands for a floating-point multiply-add can consume one to two orders of magnitude more energy than performing the operation.[11] Moreover, current designs frequently either seek to minimize worst-case performance to maximize generality or sacrifice programmability to maximize best-case performance. Future memory-systems must seek energy efficiency through specialization (e.g., through compression and support for streaming data) while simplifying programmability (e.g., by extending coherence and virtual memory to accelerators when needed). Such mechanisms have the potential to reduce energy demands for a broad range of systems, from always-on smart sensors to data centers processing big data.

**Exploiting (Re-)configurable Logic Structures.** The increasing complexity of silicon process technologies has driven NRE costs to prohibitive levels, making full-custom accelerators infeasible for all but the highest-volume applications. Current reconfigurable logic platforms (e.g., FPGAs) drive down these fixed costs, but incur undesirable energy and performance overheads due to their fine-grain reconfigurability (e.g., lookup tables and switch boxes). Research in future accelerators will improve energy efficiency using coarser-grain semi-programmable building blocks (reducing internal inefficiencies) and packet-based interconnection (making more efficient use of expensive wires). Additional efficiencies will come from emerging 3D technologies such as silicon interposers, which allow limited customization (e.g., top-level interconnect) to configure systems at moderate cost. Such techniques coupled with better synthesis tools can reduce NRE costs, thereby enabling rapid development and deployment of accelerators in a broad range of critical applications.

---

[11] Steve Keckler, "Life After Dennard and How I Learned to Love the Picojoule," Keynote at Micro 2011.

The past scaling of processor performance has driven advances both within computing and in society at large. Continued scaling of performance will be largely limited by the improvements in energy efficiency made possible by reconfigurable and specialized hardware. Research that facilitates the design of reconfigurable and special-purpose processors will enable corporations, researchers, and governments to quickly and affordably focus enormous computing power on critical problems.

## 2.3. Technology Impacts on Architecture

Application demands for improved performance, power and energy efficiency, and reliability drive continued investment in technology development. As standard CMOS reaches fundamental scaling limits, the search continues for replacement circuit technologies (e.g., sub/near-threshold CMOS, QWFETs, TFETs, and QCAs) that have a winning combination of density, speed, power consumption, and reliability.  Non-volatile storage (i.e., flash memory) has already starting to replace rotating disks in many ICT systems, but comes with its own design challenges (e.g., limited write endurance).  Other emerging non-volatile storage technologies (e.g., STT-RAM, PCRAM, and memristor) promise to disrupt the current design dichotomy between volatile memory and non-volatile, long-term storage.  3D integration uses die stacking to permit scaling in a new dimension, but substantial technology and electronic design automation (EDA) challenges remain. Photonic interconnects can be exploited among or even on chips.

In addition to existing efforts to develop new ICT technologies (e.g., through NSF's MRSECs), significant architectural advancements—and thus significant investments—are needed to exploit these technologies.

**Rethinking the Memory/Storage Stack.** Emerging technologies provide new opportunities to address the massive online storage and processing requirements of "big data" applications. Emerging non-volatile memory technologies promise much greater storage density and power efficiency, yet require re-architecting memory and storage systems to address the device capabilities (e.g., longer, asymmetric, or variable latency, as well as device wear out).

**Design Automation Challenges.**  New technologies drive new designs in circuits, functional units, microarchitectures, and systems.  Such new approaches also require investment in new EDA tools, particularly for mixed-signal and 3D designs.  New *design* tools must be tailored to the new technologies:  functional synthesis, logic synthesis, layout tools, and so on.  Furthermore, heterogeneous computing greatly taxes our ability to perform pre-RTL system modeling, particularly as the diversity of architectures and accelerators explodes. Hence, new *verification*, *analysis* and *simulation* tools will be needed:  tools for verifying correct operation, performance, power and energy consumption, reliability (e.g., susceptibility to soft error and aging effects), and security (avoiding power "footprints," providing architectural support for information flow tracking).

**3D Integration.** Die stacking promises lower latency, higher bandwidth, and other benefits, but brings many new EDA, design, and technology challenges. Computer architects have started this investigation with the stacking of DRAM memories on top of cores, but future designs will go much further to encompass stacking of nonvolatile memories, of custom computational components realized with a non-compatible technology, of new interconnection technologies (e.g., photonics), of sensors and the analog components that go with them, of RF and other analog components, of energy providers and cooling systems (e.g., MEMs energy harvesting devices, solar cells, and microfluidic cooling). To realize the promise of the new technologies being developed by nano-materials and nano-structures researchers, further investment is needed so that computer architects can turn these nanotechnology circuits into ICT systems.

## 2.4. Cross-Cutting Issues and Interfaces

As computers permeate more aspects of everyday life, making a computer system *better* means much more than being *faster* or more *energy efficient*. Applications need architectural support to ensure data security and privacy, to tolerate faults from increasingly unreliable transistors, and to enhance programmability, verifiability and portability. Achieving these cross-cutting design goals—nicknamed the "Ilities"—requires a fundamental rethinking of long-stable interfaces that were defined under extremely different application requirements and technological constraints.

**Security, Programmability, Reliability, Verifiability and Beyond.**

Applications increasingly demand a richer set of design "Ilities" at the same time that energy constraints make them more expensive to provide. Fortunately, the confluence of new system architectures and new technologies creates a rare inflection point, opening the door to allow architects to develop fundamentally more energy-efficient support for these design goals.

**Verifiability and Reliability.** Ensuring that hardware and software operate reliably is more important than ever; for implantable medical devices, it is (literally) vital. At the same time, CMOS scaling trends lead to less-reliable circuits and complex, heterogeneous architectures threaten to create a "Verification Wall". Future system architectures must be designed to facilitate hardware and software verification; for example, using co-processors to check end-to-end software invariants. Current highly-redundant approaches are not energy efficient; we recommend research in lower-overhead approaches that employ dynamic (hardware) checking of invariants supplied by software. In general, we must architect ways of continuously monitoring system health—both hardware and software—and applying contingency actions. Finally, for mission-critical scenarios (including medical devices), architects must rethink designs to allow for failsafe operation.

**Security and Privacy.** Architectural support for security dates back decades, to paging, segmentation and protection rings. Recent extensions help to prevent buffer overflow attacks, accelerate cryptographic operations, and isolate virtual machines. However, it is time to rethink security and privacy from the ground up and define architectural interfaces that enable hardware to act as the "root of trust", efficiently supporting secure services. Such services include information flow tracking (reducing side-channel attacks) and efficient enforcement of richer information access rules (increasing privacy). Support for tamper-proof memory and copy-protection are likewise crucial topics. Finally, since security and privacy are intrinsically connected to reliable/correct operation, research in these areas dovetails well.

**Improving Programmability**. Programmability refers to all aspects of producing software that reaches targets for performance, power, reliability, and security, with reasonable levels of design and maintenance effort. The past decades have largely focused on software engineering techniques—such as modularity and information hiding—to improve programmer productivity at the expense of performance and power.[12] As energy efficiency and other goals become more important, we need new techniques that cut across the layers of abstraction to eliminate unnecessary inefficiencies.

Existing efforts to improve programmability, including domain-specific languages, dynamic scripting languages, such as Python and Javascript, and others, are pieces of the puzzle. Facebook's HipHop, which dynamically compiles programs written in scripting language, shows how efficiency can be reclaimed despite such abstraction layers. In addition to software that cuts through abstraction layers, we recommend cross-cutting research in *hardware* support to improve programmability. Transactional memory (TM) is a recent example that seeks to significantly simplify parallelization and synchronization in multithreaded code. TM research has spanned all levels of the system stack, and is now entering the commercial mainstream. Additional research is required on topics like software debugging, performance bottleneck analysis, resource management and profiling, communication management, and so on.

Managing the interactions between applications also present challenges. For example, how can applications express Quality-of-Service targets and have the underlying hardware, the operating system and the virtualization layers work together to ensure them? Increasing virtualization and introspection support requires coordinated resource management across all aspects of the hardware and software stack, including computational resources, interconnect, and memory bandwidth.

## Crosscutting Interfaces

Current computer architectures define a set of interfaces that have evolved slowly for several decades. These interfaces—e.g., the Instruction Set Architecture and virtual memory—were defined when memory was at a premium, power was abundant, software infrastructures were limited, and there was little concern for security. Having stable interfaces has helped foster decades of evolutionary architectural innovations. We are now, however, at a technology crossroads, and these stable interfaces are a hindrance to many of the innovations discussed in this document.

**Better Interfaces for High-Level Information.** Current ISAs fail to provide an efficient means of capturing software-intent or conveying critical high-level information to the hardware. For example, they have no way of specifying when a program requires energy efficiency, robust security, or a desired Quality of Service (QoS) level. Instead, current hardware must try to glean some of this information on its own—such as instruction-level parallelism or repeated branch outcome sequences—at great energy expense. New, higher-level interfaces are needed to encapsulate and convey programmer and compiler knowledge to the hardware, resulting in major efficiency gains and valuable new functionality.

**Better Interfaces for Parallelism.** Developing and supporting parallel codes is a difficult task. Programmers are plagued by synchronization subtleties, deadlocks, arbitrary side effects, load imbalance and unpredictable communication, unnecessary non-determinism, confusing memory models, and performance opaqueness. We need interfaces that allow the programmer to express parallelism at a higher level, expressing localities, computation dependences and side effects, and the key sharing and communication patterns. Such an interface could enable simpler and more efficient hardware, with efficient communication and synchronization primitives that minimize data movement.

---

[12] James Larus, Spending Moore's Dividend, *Communications of the ACM*, May 2009 5(52).

**Better Interfaces for Abstracting Heterogeneity.** Supporting heterogeneous parallelism demands new interfaces. From a software perspective, applications must be programmed for several different parallelism and memory usage models; and they must be portable across different combinations of heterogeneous hardware. From a hardware perspective, we need to design specialized compute and memory subsystems. We therefore need hardware interfaces that can abstract the key computation and communication elements of these hardware possibilities. Such interfaces need to be at a high enough level to serve as a target for portable software and at a low-enough level to efficiently translate to a variety of hardware innovations.

**Better Interfaces for Orchestrating Communication.** Traditional computers and programming models have focused heavily on orchestrating computation, but increasingly it is data communication that must be orchestrated and optimized. We need interfaces that more clearly identify long-term data and program dependence relationships, allowing hardware and software schedulers to dynamically identify critical paths through the code. Without the ability to analyze, orchestrate, and optimize communication, one cannot adhere to performance, energy or QoS targets. Data management becomes even more complex when considering big-data scenarios involving data orchestration between many large systems. Current systems lack appropriate hardware-software abstractions for describing communication relationships.

**Better Interfaces for Security and Reliability.** Existing protection and reliability models do not address current application needs. We need interfaces to specify fine-grain protection boundaries among modules within a single application, to treat security as a first class property, and to specify application resilience needs or expectations. Some parts of the application may tolerate hardware faults, or may be willing to risk them to operate more power-efficiently. All these interfaces can benefit from appropriate hardware mechanisms, such as information-flow tracking, invariants generation and checking, transactional recovery blocks, reconfiguration, and approximate data types. The result will be significant efficiencies.

## 3. Closing

This white paper surveys the challenges and some promising directions for investment in computer architecture research to continue to provide better computer systems as a key enabler of the information and communication innovations that are transforming our world.

# 4. About this Document

This document was created through a distributed process during April and May 2012. Collaborative writing was supported by a distributed editor. We thank the Computing Community Consortium[13] (CCC), including Erwin Gianchandani and Ed Lazowska, for guidance, as well as Jim Larus and Jeannette Wing for valuable feedback. Researchers marked with "*" contributed prose while "**" denotes effort coordinator.

Sarita Adve, University of Illinois at Urbana-Champaign *
David H. Albonesi, Cornell University
David Brooks, Harvard
Luis Ceze, University of Washington *
Sandhya Dwarkadas, University of Rochester
Joel Emer, Intel/MIT
Babak Falsafi, EPFL
Antonio Gonzalez, Intel and UPC
Mark D. Hill, University of Wisconsin-Madison *,**
Mary Jane Irwin, Penn State University *
David Kaeli, Northeastern University *
Stephen W. Keckler, NVIDIA and The University of Texas at Austin
Christos Kozyrakis, Stanford University
Alvin Lebeck, Duke University
Milo Martin, University of Pennsylvania
José F. Martínez, Cornell University
Margaret Martonosi, Princeton University *
Kunle Olukotun, Stanford University
Mark Oskin, University of Washington
Li-Shiuan Peh, M.I.T.
Milos Prvulovic, Georgia Institute of Technology
Steven K. Reinhardt, AMD Research
Michael Schulte, AMD Research and University of Wisconsin-Madison
Simha Sethumadhavan, Columbia University
Guri Sohi, University of Wisconsin-Madison
Daniel Sorin, Duke University
Josep Torrellas, University of Illinois at Urbana Champaign *
Thomas F. Wenisch, University of Michigan *
David Wood, University of Wisconsin-Madison *
Katherine Yelick, UC Berkeley, Lawrence Berkeley National Laboratory *

---

[13] http://www.cra.org/ccc/

# Appendix A. Emerging Application Attributes

Much evidence suggests that ICT innovation is accelerating with many compelling visions moving from science fiction toward reality. Table A.1 below lists some of these visions, which include personalized medicine to target care and drugs to an individual, sophisticated social network analysis of potential terrorist threats to aid homeland security, and telepresence to reduce the greenhouse gases spent on commuting and travel. Furthermore, it is likely that many important applications have yet to emerge. How many of us predicted social networking even a few years before it became ubiquitous?

While predicted and unpredicted future applications will have varied requirements, it appears that many share features that were less common in earlier applications. Rather than center on the desktop, today the centers of innovation lie in sensors, smartphones/tablets, and the data-centers to which they connect. Emerging applications share challenging attributes, many of which arise because the applications produce data faster than can be processed within current, limited capabilities (in terms of performance, power, reliability, or their combination). Table A.2 lists some of these attributes, which include processing of vast data sets, using distributed designs, working within form-factor constraints, and reconciling rapid deployment with efficient operation.

**Table A.1: Example Emerging Applications**

| |
|---|
| **Data-centric Personalized Healthcare** - Future health systems will monitor our health 24/7, employing implantable, wearable, and ambient smart sensors. Local on-sensor analysis can improve functionality and reduce device power by reducing communication, while remote (i.e., cloud-based) systems can aggregate across time and patient populations. Such systems will allow us to query our own health data while enabling medical providers to continuously monitor patients and devise personalized therapies. New challenges will emerge in devising computing fabrics that meet performance, power, and energy constraints, in devising appropriate divisions between on-device and in-cloud functionality, as well as protecting this distributed medical information in the cloud. |
| **Computation-driven Scientific Discovery** - Today's advanced computational and visualization capabilities are increasingly enabling scientists and engineers to carry out simulation-driven experimentation and mine enormous data sets as the primary drivers of scientific discovery. Key areas that have already begun to leverage these advances are bio-simulation, proteomics, nanomaterials, and high-energy physics. Just as the scientific research community begins to leverage real data, issues with security and reproducibility become critical challenges. |
| **Human Network Analytics** - Given the advances in the Internet and personal communication technologies, individuals are interacting in new ways that few envisioned ten years ago. Human interaction through these technologies has generated significant volumes of data that can allow us to identify behaviors and new classes of connections between individuals. Advances in Network Science and Machine Learning have greatly outpaced the ability of computational platforms to effectively analyze these massive data sets. Efficient human network analysis can have a significant impact on a range of key application areas including Homeland Security, Financial Markets, and Global Health. |
| **Many More** - In addition to these three examples, numerous problems in the fields of personalized learning, telepresence, transportation, urban infrastructure, machine perception/inference and enhanced virtual reality are all pushing the limits of today's computational infrastructure. The computational demands of these problem domains span a wide range of form factors and architectures including embedded sensors, hand-held devices and entire data centers. |

### Table A.2: Attributes of Emerging Applications

The three example applications in Table A.1 share at least four key attributes that present barriers that require additional research to overcome.

**Big Data** - With the ubiquity of continuous data streams from embedded sensors and the voluminous multimedia content from new communication technologies, we are in the midst of a digital information explosion. Processing this data for health, commerce and other purposes requires efficient balancing between computation, communication, and storage. Providing sufficient on-sensor capability to filter and process data where it is generated/collected (e.g., distinguishing a nominal biometric signal from an anomaly), can be most energy-efficient, because the energy required for communication can dominate that for computation. Many streams produce data so rapidly that it is cost-prohibitive to store, and must be processed immediately. In other cases, environmental constraints and the need to aggregate data between sources impacts where we carry out these tasks. The rich tradeoffs motivate the need for hybrid architectures that can efficiently reduce data transfer while conserving energy.

**Always Online** - To protect our borders, our environments and ourselves, computational resources must be both available and ready to provide services efficiently. This level of availability places considerable demands on the underlying hardware and software to provide reliability, security and self-managing features not present on most systems. While current mainframes and medical devices strive for five 9's or 99.999% availability (all but five minutes per year), achieving this goal can cost millions of dollars. Tomorrow's solutions demand this same availability at the many levels, some where the cost is only a few dollars.

**Secure and Private** - As we increase our dependence on data, we grow more dependent on balancing computational performance with providing for available, private, and secure transactions. Information security is a national priority, and our computational systems are presently highly vulnerable to attacks. Cyber warfare is no longer a hypothetical, and numerous attacks across the Internet on government websites have already been deployed. New classes of hardware systems, architectures, firmware and operating systems are required to provide for the secure delivery of distributed information for the range of applications described above.